\documentclass[twoside,fleqn]{article}
\usepackage{espcrc2,amssymb,hyperref}

\def\openone{\leavevmode\hbox{\small1\normalsize\kern-.33em1}}

\def\meff{m_{\rm eff}}
\def\sp{s^\prime}

\def\gfive{(\gamma_5\otimes\openone)}

\def\gxone{(\openone\otimes\openone)}
\def\gonexfivemu{(\openone\otimes\xi_{5\mu})}
\def\gfivexfivemu{(\gamma_5\otimes\xi_{5\mu})}

\title{Staggered Domain Wall Fermions}
 
\author{
  George T.~Fleming
  \address{
    Physics Department,
    The Ohio State University,
     Columbus, OH 43210-1168, USA.
  } and
  Pavlos M.~Vranas
  \address{
    IBM T.J. Watson Research Center, 
    Route 134,
    Yorktown Heights, NY 10598, USA.
  }
}

\begin{document}

\pagestyle{empty}

\begin{abstract}

Staggered Domain Wall Fermions (SDWF) combine the attractive chiral properties
of staggered fermions with those of domain wall fermions.  SDWF describe
four flavors with exact U(1)$\times$U(1) flavor chiral symmetry.
An extra lattice dimension is introduced and the full SU(4)$\times$SU(4)
flavor chiral symmetry is recovered as its size is increased.  Here,
the free theory of SDWF is described and a preliminary discussion
of the interacting case is presented. SDWF may be well suited
for numerical simulation of lattice QCD thermodynamics.

\end{abstract}

\maketitle  

\section{Introduction}
\label{sec:introduction}

Numerical simulations of QCD thermodynamics, using current methods and
supercomputers do not realize the flavor chiral symmetries as well as one
would like. More specifically the study of the relevant critical phenomena
close to the finite temperature transition may be affected.  It is reasonable
to investigate alternatives to current methods especially if they combine some
of their attractive features.

Staggered fermions, and their variations, describe four flavors
(in four dimensions) and have an exact U(1)$\times$U(1) flavor chiral symmetry.
The full SU(4)$\times$SU(4) symmetry is recovered when the lattice spacing $a$
approaches zero.  Two flavor simulations of QCD are done by taking
the ``square root'' of the staggered determinant.  Because the U(1)$\times$U(1)
chiral symmetry is exact at any $a$ staggered fermions have been attractive
for studies of spontaneous chiral symmetry breaking phenomena.
On the other hand only one pion in the multiplet acts as a Goldstone boson
due to the flavor breaking.  Even the highly improved {\tt asqtad} action
may require $a^{-1} \gtrsim 8 T_c$ to get all pions lighter than the kaon mass
\cite{Bernard:2001tx}.

A few years ago new methods were developed with good chiral symmetry
properties:  please see the review articles in
\cite{Narayanan:1994gt,Creutz:1995px,Shamir:1996zx,Blum:1999ud,Neuberger:2000ry,Vranas:2001tz,Edwards:2001,Kikukawa:2001}
and references therein. One of them,
domain wall fermions (DWF), uses standard Wilson fermions but introduces
an extra dimension with $L_s$ sites and free boundary conditions.  As a result
massless surface modes develop, with one chirality bound exponentially
on one wall and the other on the other wall.  At finite $L_s$ chiral symmetry
is broken by their overlap but it is restored exponentially fast as $L_s$
is increased.  At $L_s = \infty$ (the overlap formalism
\cite{Narayanan:1994gt,Neuberger:2000ry}) the chiral symmetry is exact,
even at non zero $a$ and topological zero mode effects are reproduced.
These remarkable properties make DWF a good candidate for QCD thermodynamic
studies.  For such studies see
\cite{Fleming:2000xe,Vranas:2000dg,Chen:2001zu,Fleming:2001bk}. 

Both staggered and DWF are very close to realistic simulations
of QCD thermodynamics.  The present work is an attempt to combine
their attractive features.  SDWF are staggered fermions defined in space time
with one extra direction and free boundary conditions.  As with DWF,
surface modes develop with the plus chirality of all flavors localized
on one boundary (wall) and the other chirality on the other boundary.
By properly identifying the various flavors on the boundaries one can construct
four flavors of Dirac spinors with exact U(1)$\times$U(1) at non-zero $a$
and finite $L_s$.  The full SU(4)$\times$SU(4) symmetry is recovered
in the $L_s\to\infty$ limit even for non zero $a$.

Here, the free theory of SDWF is described and a preliminary discussion
of the interacting case is presented. At this moment it is not clear
to what degree the localization properties persist in the strong
interaction regime where lattice QCD thermodynamic simulations are
presently done.

\section{Staggered Domain Wall Fermions}
\label{sec:sdwf}

We start our construction of the free SDWF action by writing the free
staggered action in momentum space and in the hypercubic notation
of Kluberg-Stern {\it et al.} \cite{Kluberg-Stern:1983dg}
\begin{equation}
S_{\rm sf} = {\textstyle \sum_k} \overline{Q}(k) \left[
m (\openone\otimes\openone) + \not\!\!D(k)
\right] Q(k)
\end{equation}
\begin{eqnarray}
\label{eq:dslash}
\not\!\!D(k) = {\textstyle \sum_\mu} \left[
  \textstyle\frac{i}{2} \sin k_\mu (\gamma_\mu\otimes\openone)
  + b_\mu (\gamma_5\otimes\xi_{5\mu})
\right]
\end{eqnarray}
with $b_\mu = (1-\cos k_\mu)/2$.  The lattice coordinates $x_\mu$ are
decomposed into the coordinates of $2^d$ hypercubes $y_\mu$ and coordinates
of sites within a given hypercube $A_\mu = ( x_\mu \bmod 2 ) \in \{ 0, 1 \}$
so that $x_\mu = 2 y_\mu + A_\mu$.  The momenta $k_\mu$ are reciprocal
to the hypercubes $y_\mu$.

To complete the construction of the free SDWF action, we introduce an extra
dimension in the $s$ direction with free boundaries, sum $S_{\rm sf}$
over the $s$-slices and add terms to the action that project on
$(\gamma_5\otimes\openone)$ chiralities
\begin{equation}
\label{eq:sdwf_action}
S_{\rm sdwf} = {\textstyle \sum_s} \left[
  S_{\rm sf} + {\textstyle \sum_k} \overline{Q}(k,s) D_5 Q(k,s)
\right]
\end{equation}
\begin{equation}
D_5(s,\sp) = \textstyle\frac{1}{a_5} \left[
  P_+ \delta_{s+1,\sp} + P_- \delta_{s-1,\sp} - \delta_{s,\sp}
\right].
\end{equation}
Explicitly, the chiral projectors are
$P_\pm \equiv ((\openone\pm\gamma_5)\otimes\openone)/2$.
Even though $D_5$ breaks the U(1)$\times$U(1) symmetry generated
by $(\gamma_5\otimes\xi_5)$, we will show in section \ref{sec:symmetries}
that an extended symmetry exists to protect mass terms like
$m (\openone\otimes\openone) \delta_{s,\sp}$
from additive renormalizations.  Thus, we find it natural to set $m \to
\frac{1}{a_5}$ and drop these terms from the action.  However, it is important
to keep in mind that at higher energy scales, fermions propagate like
{\it massive} staggered fermions in $d$ dimensions.

Fully chiral flavor symmetric modes should exist at lower energy scales
in the SDWF theory if we can find normalizable states that satisfy
the following zero mode equation
\begin{eqnarray}
\label{eq:zero_mode_problem}
\lefteqn{
  {\textstyle \sum_{\sp}} \left\{
    \textstyle\frac{1}{a_5} \left[
      P_+ \delta_{s+1,\sp} + P_- \delta_{s-1,\sp}
    \right]
  \right.
} \nonumber \\*
& + & \left.
  {\textstyle \sum_\mu} (\gamma_5\otimes\xi_{5\mu})
  b_\mu \delta_{s,\sp}
\right\} \Phi(k,\sp) = 0 .
\end{eqnarray}
From this equation, it is easy to see that the $P_\pm$ projectors
in the $s$-dependent part should commute with the flavor breaking part
so that each may be simultaneously diagonalized.  This constraint alone
effectively restricts the allowed projectors to the ones we have chosen.
We define an additional set of operators to project out
specific flavor components as well:
$ P_{+\pm} \equiv (P_+\otimes(\openone\pm\xi_5))/2$.
The solution is separable and $\phi(s)$ is the $s$-dependent part.
We write it block notation using the new projectors:
\begin{equation}
\label{eq:block_vector}
\phi(s)^T = \left(
\phi^T_{++}, \phi^T_{+-}, \phi^T_{-+}, \phi^T_{--} 
\right)
\end{equation}
where $\phi_{-+}(s) = P_{-+} \phi(s)$, etc.  In this notation,
we can write
\begin{equation}
\label{eq:B_matrix}
{\textstyle \sum_\mu} (\gamma_5\otimes\xi_{5\mu}) b_\mu = \left(
  \begin{array}{rrrr}
             & B &           &    \\
  -B^\dagger &   &           &    \\
             &   &           & -B \\
             &   & B^\dagger &
  \end{array}
\right) .
\end{equation}
The solutions to (\ref{eq:zero_mode_problem}) (after iterating one time) are
\begin{equation}\begin{array}{rcl}
\label{eq:zero_mode_solution}
\phi_{\pm+}(s\pm2) & = & - a_5^2 B B^\dagger \phi_{\pm+}(s) \\
\phi_{\pm-}(s\pm2) & = & - a_5^2 B^\dagger B \phi_{\pm-}(s) .
\end{array}\end{equation}
Solving the equations relating nearest neighbor $s$ sites is more complicated
because the flavor components mix and will be discussed elsewhere
\cite{Fleming:2001}.

For free fermions, $[B,B^\dagger]=0$ and $a_5^2 B B^\dagger$,
$a_5^2 B^\dagger B$ are both proportional to the identity with eigenvalue
\begin{equation}
\lambda(a_5^2 B B^\dagger) = \lambda(a_5^2 B^\dagger B)
= a_5^2 {\textstyle \sum_\mu} b_\mu^2 .
\end{equation}
If we require that $a_5^2 {\textstyle \sum_\mu} b_\mu^2 < 1$ then
for a semi-infinite $s$ direction, $s$$\ge$0, only $\phi_{+\pm}$
is normalizable, while $\phi_{-\pm}$ is not. However, this is not enough
to ensure that the doubler modes are not present.  One must further require
that the above condition excludes momenta with components larger or
equal to $\pi$.  This is satisfied provided that $a_5$$\ge$1.  If one wanted
to further restrict the momenta to $\pi/2$ then the requirement is
that $a_5$$\ge$2.

Most of the recent work in the overlap formalism is related to taking
the $a_5$$\to$0 limit.  Clearly, additional terms must be added to our
SDWF action to maintain the cutoff of doubler momenta as $a_5$$\to$0.
Possible terms are currently under study.  In any construction of a staggered
overlap Hamiltonian, it will be important to demonstrate that naive fermions
are not recovered in the chiral limit and that the U(1)$\times$U(1) symmetry
remains exact.  Of course, it is always possible to use the overlap formalism
directly with finite $a_5$.

When adding interactions to a free staggered action, it is usually
a good idea to first transcribe the free action in terms of single component
per site fields in position space.  The unitarily equivalent hypercubic basis
of Daniel and Sheard \cite{Daniel:1988aa} is ideally suited for this purpose.
In this basis, fermion bilinears are written
$\overline{\chi}_A(y) (\overline{\gamma_S\otimes\xi_F})_{AB} \chi_B(y)$,
where the fermion fields $\chi_A(k)$ are directly the single component fields
on the corners of the hypercube $y$: $\chi(x) = \chi(2y+A) = \chi_A(y)$.
The components of the spin-flavor matrices are given by
\begin{equation}
\left(\overline{\gamma_S\otimes\xi_F}\right)_{AB}
= \textstyle\frac{1}{2^{d/2}} {\rm tr}
[ \gamma_A^\dagger \gamma_S \gamma_B \gamma_F^\dagger ]
\in \{ 0, \pm 1 \}
\end{equation}
where $S$ and $F$ are $d$-dimensional binary vectors like $A$ and $B$
and $\gamma_S \equiv \gamma_1^{S_1} \times \cdots \times \gamma_d^{S_d}$.

The staggered $\not\!\!D$ is the usual one
\begin{equation}
\label{eq:dslash_single_comp}
\not\!\!D = {\textstyle \sum_\mu} (-1)^{\eta_\mu(x)} \left[
  \delta_{x+\hat\mu,x^\prime} - \delta_{x,x^\prime+\hat\mu}
\right]
\end{equation}
where $\eta_\mu(x) = \sum_{\nu<\mu} x_\nu$.  For the projection terms in $D_5$
proportional to $(\overline{\gamma_5\otimes\openone})$, we have
\begin{eqnarray}
\lefteqn{\overline{\chi}(y,s)(\overline{\gamma_5\otimes\openone})\chi(y,s\pm1)}
\nonumber \\*
& = & (-1)^{\varphi(x)} \overline{\chi}(x,s) \chi(\overline{x},s\pm1)
\end{eqnarray}
where $\varphi(x) = d/2 + \sum_{\mu=1}^{d/2} x_{2\mu-1}$ and $\overline{x}$
is the opposite corner of the hypercube:
$\overline{x}_\mu = x_\mu + 1-2(x_\mu \bmod 2)\ \forall\ \mu$.

\section{Symmetries}
\label{sec:symmetries}

When constructing the SDWF action, it is important to preserve
all the symmetries of the massless staggered action \cite{Jolicoeur:1986ek}.
Of course, adding any new terms to the staggered action will likely break some
of those symmetries, so we have to find new symmetries that involve
the extra dimension.  However, we will still have to show that these
new symmetries in the space with an extra direction are maintained
by the action that has all fields integrated out except the
``light'' ones at the boundaries. We present here the symmetry
transformations in the hypercubic notation for the space with
an extra dimension.

{\it Rotations by $\pi/2$}. These rotations are in planes perpendicular
to the extra dimension and the transformations are the same as the original
staggered ones.

{\it $\mu$-parity}.  These transformations reflect the $d-1$ spatial axes
perpendicular to the spatial axis in the $\hat\mu$ direction.  $D_5$ is not
invariant under this symmetry unless we also reflect the $s$ direction as well.
If we define the reflection operator ${\cal R}_{s,\sp}
\equiv \delta_{L_s-1-s,\sp}$, the transformation is
\begin{equation} \begin{array}{rcl}
Q(y,s) & \to & (\gamma_\mu\otimes\xi_5) {\cal R}_{s,\sp}\ Q(y,\sp) , \\*
\overline{Q}(y,s) & \to &
  \overline{Q}(y,\sp)\ {\cal R}_{\sp,s} (\gamma_\mu\otimes\xi_5) .
\end{array} \end{equation}

{\it Shift by one lattice spacing}.  By inspection, one can see that
(\ref{eq:dslash_single_comp}) is invariant under shifts by one lattice spacing
in any spatial direction $\mu$.  Because of the additional structure imposed
on the lattice by the hypercubic formulation, the symmetry transformation,
while still valid, is complicated.  For SDWF, there is an added complication
that some parts of the transformation require a reflection
in the $s$ direction (with $s$ indices suppressed):
\begin{eqnarray}
\lefteqn{ Q(y) \to {\textstyle \frac{1}{2}} \left[
  (\openone\otimes\xi_\mu) - (\gamma_{\mu5}\otimes\xi_5) {\cal R}
\right] Q(y) } \nonumber \\*
& & + {\textstyle \frac{1}{2}} \left[
  (\openone\otimes\xi_\mu) + (\gamma_{\mu5}\otimes\xi_5) {\cal R}
\right] Q(y+\hat\mu) , \nonumber \\*
\lefteqn{ \overline{Q}(y) \to \overline{Q}(y) {\textstyle \frac{1}{2}}
\left[
  (\openone\otimes\xi_\mu) - {\cal R} (\gamma_{5\mu}\otimes\xi_5)
\right] } \nonumber \\*
& & + \overline{Q}(y+\hat\mu) {\textstyle \frac{1}{2}} \left[
  (\openone\otimes\xi_\mu) + {\cal R} (\gamma_{5\mu}\otimes\xi_5)
\right] .
\end{eqnarray}

U(1)$_e$$\times$U(1)$_o$ {\it chiral rotations}.  The residual chiral symmetry
of staggered fermions involves making separate chiral rotations on even and
odd sites.  Terms in $D_5$ are not invariant under these rotations unless
we extend the notion of even and odd, {\it including the extra dimension}.
First, we define the operator
${\cal S}_{s,\sp} \equiv (-1)^s  \delta_{s,\sp}$ and then the
extended even/odd projection operators
\begin{eqnarray}
P_e & = & {\textstyle \frac{1}{2}} \left[
  (\openone\otimes\openone) + {\cal S} (\gamma_5\otimes\xi_5)
\right] , \nonumber \\*
P_o & = & {\textstyle \frac{1}{2}} \left[
  (\openone\otimes\openone) - {\cal S} (\gamma_5\otimes\xi_5)
\right] .
\end{eqnarray}
Using these projection operators the chiral transformation is
\begin{eqnarray}
Q(y) & \to & \left( e^{i\theta_e} P_e + e^{i\theta_o} P_o \right)\ Q(y) ,
\nonumber \\*
\overline{Q}(y) & \to & \overline{Q}(y)\ 
\left( e^{-i\theta_o} P_e + e^{i\theta_e} P_o \right) .
\end{eqnarray}
%

\section{Flavors of SDWF}
\label{sec:flavors}

From section \ref{sec:sdwf} one sees that for a finite extra direction
with $L_s$ sites the $P_+$ components of all flavors are localized
around $s$=0 while the $P_-$ components are localized around $s$=$L_s$$-$1.
From section \ref{sec:symmetries}, however, we mentioned
the importance of constructing an effective $d$-dimensional field $q$
from the surface states consistent with all the SDWF symmetries, particularly
the U(1)$_e$$\times$U(1)$_o$ chiral symmetry.  For example, to project flavor 
components with $P_{++}$, one should choose $s$ near zero.  If $s$=0 is chosen,
$P_{++} q(y) = P_{++} Q(y,0)$, then these components also belong
to the $P_e$ part of the fermion field.  Therefore, to project flavor
components with $P_{-+}$ one is not only restricted to choose $s$
near $L_s$$-$1 but also choose $s$ so these components belong to the $P_o$ part
of the fermion field.  Then, components $P_{\pm+} q$ will not mix even
for finite $L_s$ because of the even/odd symmetry.  In this example, one would
like to pick $P_{-+} q(y) = P_{-+} Q(y,s)$ with $s$ being even
and near $L_s$$-$1.  So, if $L_s$ is odd then $s$=$L_s$$-$1 is a good choice.
However, if $L_s$ is even, then one should choose $s$=$L_s$$-$2 instead.

\begin{equation}
\begin{array}{c}
\left[ 
\begin{array}{c}
\left(
\begin{array}{c} 
{\rm X} \\ 
x \\
x \\
x 
\end{array}
\right) \\ \\
\left(
\begin{array}{c} 
x \\ 
{\rm X} \\
x \\
x 
\end{array}
\right) \\ \\
\vdots \\ \\
\left(
\begin{array}{c} 
x \\ 
x \\
{\rm X} \\
x 
\end{array}
\right) \\ \\
\left(
\begin{array}{c} 
x \\ 
x \\
x \\
{\rm X} 
\end{array}
\right)
\end{array}
\right] \\ \\
Q
\end{array}
\!\!\!\!=\!\!\!\!
\begin{array}{c}
\left[ 
\begin{array}{c}
\left(
\begin{array}{c} 
{\rm X} \\ 
0 \\
0 \\
x 
\end{array}
\right) \\ \\
\left(
\begin{array}{c} 
0 \\ 
{\rm X} \\
x \\
0 
\end{array}
\right) \\ \\
\vdots \\ \\
\left(
\begin{array}{c} 
x \\ 
0 \\
0 \\
x 
\end{array}
\right) \\ \\
\left(
\begin{array}{c} 
0 \\ 
x \\
x \\
0 
\end{array}
\right)
\end{array}
\right] \\ \\
P_e Q
\end{array}
\!\!\!\!+\!\!\!\!
\begin{array}{c}
\left[ 
\begin{array}{c}
\left(
\begin{array}{c} 
0 \\ 
x \\
x \\
0 
\end{array}
\right) \\ \\
\left(
\begin{array}{c} 
x \\ 
0 \\
0 \\
x 
\end{array}
\right) \\ \\
\vdots \\ \\
\left(
\begin{array}{c} 
0 \\ 
x \\
{\rm X} \\
0 
\end{array}
\right) \\ \\
\left(
\begin{array}{c} 
x \\ 
0 \\
0 \\
{\rm X} 
\end{array}
\right)
\end{array}
\right] \\ \\
P_o Q
\end{array}
\label{eq:flavors_ls_even}
\end{equation}

Using the block notation of (\ref{eq:block_vector}), an example for even $L_s$
is sketched in (\ref{eq:flavors_ls_even}).  In this equation $Q$($s$=0) is
at the top and $Q$($s$=$L_s$$-$1) is at the bottom.  The capital letters
denote one of the ``correct'' choices.  On the other hand if $L_s$ is odd,
say $3$ then
\begin{equation}
\begin{array}{c}
\left[ 
\begin{array}{c}
\left(
\begin{array}{c} 
{\rm X} \\ 
{\rm X} \\
x \\
x 
\end{array}
\right) \\ \\
\left(
\begin{array}{c} 
x \\ 
x \\
x \\
x 
\end{array}
\right) \\ \\
\left(
\begin{array}{c} 
x \\ 
x \\
{\rm X} \\
{\rm X}
\end{array}
\right)
\end{array}
\right] \\ \\
Q
\end{array}
\!\!\!\!=\!\!\!\!
\begin{array}{c}
\left[ 
\begin{array}{c}
\left(
\begin{array}{c} 
{\rm X} \\ 
0 \\
0 \\
x 
\end{array}
\right) \\ \\
\left(
\begin{array}{c} 
0 \\ 
x \\
x \\
0 
\end{array}
\right) \\ \\
\left(
\begin{array}{c} 
x \\ 
0 \\
0 \\
{\rm X} 
\end{array}
\right)
\end{array}
\right] \\ \\
P_e Q
\end{array}
\!\!\!\!+\!\!\!\!
\begin{array}{c}
\left[ 
\begin{array}{c}
\left(
\begin{array}{c} 
0 \\ 
{\rm X} \\
x \\
0 
\end{array}
\right) \\ \\
\left(
\begin{array}{c} 
x \\ 
0 \\
0 \\
x 
\end{array}
\right) \\ \\
\left(
\begin{array}{c} 
0 \\ 
x \\
{\rm X} \\
0 
\end{array}
\right)
\end{array}
\right] \\ \\
P_o Q
\end{array}
\label{flavors_ls_odd}
\end{equation}
We note that other choices for selecting flavor components near the boundaries
are certainly possible.  We also note that the even $L_s$ choice
in (\ref{eq:flavors_ls_even}) does not transform simply under the ${\cal R}$
reflections of section \ref{sec:symmetries} and requires modification
to make it consistent with the other staggered symmetries.  This will be
discussed in more detail elsewhere \cite{Fleming:2001}.

\section{The SDWF propagator}
\label{sec:propagator}

The SDWF propagator for the Dirac matrix given in section \ref{sec:sdwf}
is presented here in a general form.  The detailed form will be presented
elsewhere \cite{Fleming:2001}.  A degenerate mass term that explicitly mixes
chiralities is added.  For odd $L_s$ it has the form
\begin{eqnarray}
\lefteqn{M_{s,\sp} = -i m_f \delta_{s,L_s-1} \delta_{0,\sp} P_+}
\nonumber \\*
&& + i m_f \delta_{s,0} \delta_{L_s-1,\sp} P_- .
\label{mass_term}
\end{eqnarray}
For even $L_s$ it has a form according to the discussion in section
\ref{sec:flavors}.  The propagator in momentum space has the general form:
\begin{eqnarray}
\label{prop}
\lefteqn{D^{-1}(s,\sp) =} \\*
&& [ G_1 + m_f G_2] \epsilon(s - \sp) + G_3 \epsilon(s - \sp - 1) \nonumber
\end{eqnarray}
where $G_1$, $G_2$ and $G_3$ are functions of momenta and are proportional
to the identity in their flavor indices.  Also, $G_1$ anti-commutes
with $\gfive$ while $G_2$ and $G_3$ commute with $\gfive$.  The flavor mixing
is in the function $\epsilon$:
\begin{equation} \begin{array}{rclr} \displaystyle
\epsilon(x) & = & \gxone , & (x\ {\rm even}) \\*
\epsilon(x) & = &
{\textstyle\sum_\mu \gonexfivemu b_\mu \over \textstyle |b|}, &
(x\ {\rm odd})
\label{epsilon}
\end{array} \end{equation}
where $|b| = \sqrt{\sum_\mu b_\mu^2}$.  For $s-\sp$ even and $m_f$=0
the propagator anti-commutes with $\gfive$ and has no flavor mixing
except for the last term in (\ref{prop}).  In this term $\epsilon({\rm odd})$
breaks flavor in exactly the same way as free staggered fermions.
An exact U(1)$\times$U(1) symmetry is maintained.  The matrix coefficient
$G_3$ vanishes exponentially fast with $L_s$ for $s$, $\sp$ near opposing
boundaries and therefore as $L_s \to \infty$ with $m_f$=0 the propagator
anti-commutes with $\gfive$ and has no flavor mixing provided $s-\sp$ is even.
This is in accordance with the discussion in section \ref{sec:flavors}.

Finally, if we add to $b_\mu$ a constant term $1/a_5 - m_0$:
$b_\mu = (1 - \cos k_\mu + 1/a_5 - m_0)/2$, the effective mass $\meff$
has exactly the same form as in Wilson DWF but the decay coefficient is now
in terms of $|b|$ instead of $b = \sum_\mu [ 1 - cos(p_\mu)] + 1/a_5 - m_0$.
In this case, the localization condition for $m_0$ is the same as in DWF
and the $a_5$$\to$0 limit can be taken the same way as in DWF.
Furthermore, this term may be needed to cancel any renormalization
of the flavor breaking term.  This term would appear in the action
as $(1/a_5 - m_0)/2 \sum_\mu \overline{Q} (\gamma_5\otimes\xi_{5\mu}) Q$,
a term that is not invariant under shifts by a single lattice spacing.
However, such a term, or a generalization of it, might be needed
for the above reasons.

\section{The SDWF transfer matrix}
\label{sec:tmatrix}

We can use the technique of Neuberger \cite{Neuberger:1998bg} to rewrite
the free SDWF determinant in a form that allows us to quickly identify
the transfer matrix.  We use the same basis as in (\ref{eq:B_matrix}).
After interchanging various rows and columns of the SDWF matrix,
the determinant is equivalent to the determinant of the matrix
\begin{equation} \left( \begin{array}{cccc}
  \alpha_0 &        &                & \beta_0        \\
  \beta_1  & \ddots &                &                \\
           & \ddots & \alpha_{L_s-2} &                \\
           &        & \beta_{L_s-1}  & \alpha_{L_s-1}
\end{array} \right) \end{equation}
where all of the $\alpha_s$ and $\beta_s$ are the block triangular matrices
\begin{equation}
\alpha_s = \left( \begin{array}{cc}
  {\cal B}  & 0     \\
  -{\cal C} & 1/a_5
\end{array} \right) ,\ 
\beta_s = \left( \begin{array}{cc}
  1/a_5 & {\cal C}^\dagger \\
  0     & -{\cal B} 
\end{array} \right) .
\end{equation}
For $\alpha_{L_s-1}$ and $\beta_0$, $1/a_5$ is replaced with $-\mu/a_5$
so $\mu$ is a parameter that controls the boundary conditions:
$\mu = \pm 1$ for (anti)periodic and $\mu=0$ for free.  The definition
of ${\cal B}$ follows from (\ref{eq:B_matrix})
\begin{equation}
{\cal B} = \left( \begin{array}{cc}
  0          & B \\
  -B^\dagger & 0 
\end{array} \right)
\end{equation}
and the definition of ${\cal C}$ follows from (\ref{eq:dslash})
\begin{equation}
\sum_\mu \frac{i}{2} \sin k_\mu (\gamma_\mu\otimes\openone) = \left(
\begin{array}{cc}
  0                & -{\cal C} \\
  {\cal C}^\dagger & 0
\end{array} \right)
\end{equation}
In this notation, following Neuberger's construction leads
to the free SDWF transfer matrix
\begin{equation}
T = \left( \begin{array}{cc}
  {\cal B}^{-1}/a_5               & -{\cal B}^{-1} {\cal C} \\
  -{\cal C}^\dagger {\cal B}^{-1} &
    a_5 ({\cal C}^\dagger {\cal B}^{-1} {\cal C} - {\cal B})
\end{array} \right) .
\end{equation}

Since we chose to set the staggered mass $m \to \frac{1}{a_5}$
in (\ref{eq:sdwf_action}), then ${\cal B}$ is strictly anti-Hermitian,
so $T$ is as well (This is different from Wilson DWF and gives
some idea why solving the zero mode problem in (\ref{eq:zero_mode_solution})
simplifies when solving for the field two sites away).
In this case taking the $a_5$$\to$0 limit in order to identify the Hamiltonian
does not make sense since the doublers are re-introduced
(see section \ref{sec:sdwf}).  On the other hand if we do not set
the staggered mass $m \to \frac{1}{a_5}$ then the $a_5 \to 0$ limit does not
re-introduce the doublers and can be taken.  However this term breaks
the exact U(1)$\times$U(1) symmetry.  An alternative is to add
an $\frac{1}{a_5}-m_0$ term as in section \ref{sec:propagator}.
This does not break the U(1)$\times$U(1) symmetry,
allows the $a_5$$\to$0 limit to be taken without reintroducing the doublers,
but breaks the shift by one lattice spacing symmetry.

\section{Alternative actions}
\label{sec:alt}

The SDWF action considered here is not unique. Better actions
may be constructed using improved fields in the same spirit
as with staggered fermions (see \cite{Luo:1997vt,Lee:1999zx}
and references therein).  But even in the spin/flavor basis considered here
one could add the domain wall part in slightly different ways.  For example
one could have added to the standard staggered action the exact same term
as the one added in Wilson DWF but multiplied by $\sum_\mu \gfivexfivemu$.

\section{Conclusions}
\label{sec:conclusions}

Staggered domain wall fermions (SDWF) have been constructed for the free theory.
They describe four flavors with exact U(1)$\times$U(1) flavor chiral symmetry. 
The full SU(4)$\times$SU(4) flavor chiral symmetry is recovered as the size
of the extra dimension is increased.  The addition of interactions for QCD
is straight forward but it remains to be seen how well the SDWF properties
are maintained at the stronger couplings where numerical simulations are done.
It is hoped that SDWF may be useful for numerical simulations
of QCD thermodynamics.

\end{document}